# On the Implementation of Fixed-point Exponential Function for Machine Learning and Signal Processing Accelerators

M. Chandra

*Abstract*— **The natural exponential function is widely used in modeling many engineering and scientific systems. It is also an integral part of many neural network activation function such as sigmoid, tanh, ELU, RBF etc. Dedicated hardware accelerator and processors are designed for faster execution of such applications. Such accelerators can immensely benefit from an optimal implementation of exponential function. This can be achieved for most applications with the knowledge that the exponential function for a negative domain ($\mathscr{R}^-$) is more widely used than the positive domain ($\mathscr{R}^+$). This paper presents an optimized implementation of exponential function for variable precision fixed point negative input. The implementation presented here significantly reduces the number of multipliers and adders. This is further optimized using mixed world-length implementation for the series expansion. The reduction in area and power consumption is more than 30% and 50% respectively over previous equivalent method.**

## I. INTRODUCTION

The natural exponential function is widely used in modeling many engineering and scientific systems such as:

**Decay:** $y = y_0 e^{-\alpha t}$ $\qquad t > 0$

**Gaussian Distribution:** $y = y_0 e^{-\frac{(x-\mu)^2}{2\sigma^2}}$

It is also a building block for many non-linear activation functions used in machine learning application [1]; such as:

**Sigmoid**: $y = \frac{1}{1+e^{-x}} = \begin{cases} \frac{1}{1+e^{-|x|}} & x \geq 0 \\ 1 - \frac{1}{1+e^{-|x|}} & x < 0 \end{cases}$

**tanh**: $y = \frac{e^x - e^{-x}}{e^x + e^{-x}} = \begin{cases} \frac{1-e^{-2|x|}}{1+e^{-2|x|}} & x \geq 0 \\ \frac{e^{-2|x|}-1}{e^{-2|x|}+1} & x < 0 \end{cases}$

**ELU**: $y = \begin{cases} x & x \geq 0 \\ \propto (e^x - 1) & x < 0 \end{cases} = \begin{cases} x & x \geq 0 \\ \propto (e^{-|x|} - 1) & x < 0 \end{cases}$

From above equations, it is clear that the exponential function for a negative real domain ($\mathscr{R}^-$) is more widely used than the positive domain ($\mathscr{R}^+$). i.e., we actually use $e^{-|x|}$ more often than $e^{-x}$. This is important as the function $e^{-|x|}$ provides more opportunities to optimize compared to the implementation of $e^x$ or $e^{-x}$ i.e., exponential function for the full domain of real

numbers. It has nice properties, such as fast saturation, constrained range ($e^{-|x|} \leq 1$), which help in the optimization. Since, most application either satisfy the condition of negative domain requirement or can be manipulated mathematically to satisfy it; such an implementation is worth exploring. In this case, the hardware accelerators can be optimally designed with the support of $e^{-x}$ and then the application can be modified if needed to satisfy the constraint.

Look up tables and polynomial approximation have been used for implementing exponential and many non-linear functions [2,3]. Either of these two methods used in standalone suffers from drawbacks as the hardware requirement increases significantly for higher accuracy; so, often a hybrid approach is used. There are many variations of this approach which are discussed in detail in next paragraphs. Other than these two methods, CORDIC [4] algorithm has also been used for computation of exponential functions [5]. CORDIC is an iterative algorithm and converges slowly. Pouyan et al. have used parabolic synthesis [6] method for fast approximation of exponential function. Parabolic synthesis method approximates a functions with multiplication of multiple sub-functions. These sub-functions are parabolic functions and can be computed in parallel. In their implementation, the parallel computation improved the speed of the operation. This implementation requires four sub-functions and requires large number of multipliers.

The LUT, series expansion or hybrid method remains the most efficient and popular method for exponential implementation. Different flavors of these methods have been proposed in literature. Nilsson et al. [3] implemented the exponential function using 6th order Taylor series approximation around x=0.5. They optimized the circuit to remove an adder and a multiplier. The circuit supports 15-bit positive fractional numbers as the input.

Partzsch et al. present a hybrid approach where LUT and series approximations are combined [7]. It supports signed real numbers in s16.15 format at the input. The input is divided in three parts, integer part, higher fractional ($\geq 2^{-6}$) and lower fractional ($< 2^{-6}$). Final exponential is computed as multiplication of the exponential of each of these parts. LUTs are used for first two parts and 4th order Taylor series approximation is used for the last one. Saturation is used if the

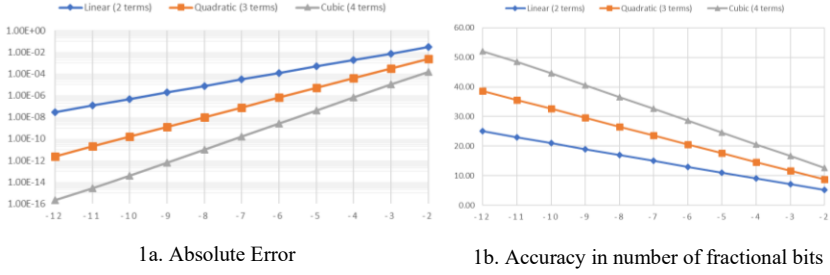

1a. Absolute Error    1b. Accuracy in number of fractional bits

Fig. 1. Maximum absolute error and accuracy (Y-axis) plotted as function of the range of the input as power of 2 (X-axis) for different number of terms in approximation (linear (2 terms), quadratic (3 terms) etc.). For this plot, accuracy is defined as how many (most significant) bits are always correctly computed using series expansion.

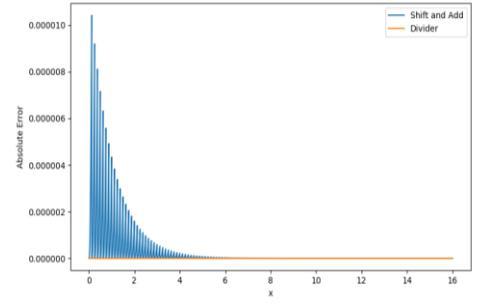

Fig. 2. Absolute error comparison on original and hardware friendly coefficient of cubic term

output is more than $2^{16}$ or less than $2^{-15}$. This corresponds to input values of 11.1 and -10.4 respective. The saturation helps in reducing the size of the LUTs. The integer and fractional LUTs have 23 and 64 entries respectively. For ease of hardware implementation the coefficients of the Taylor series are approximated as given below:

$$f(q) = 1 + q + \frac{1}{2} \times q^2 + C_3 \times q^3 + C_4 \times q^4 \qquad (1)$$

Where $C_3 = 0.1666259765625$, and $C_4 = 0.04296875$. These have been approximated so that multipliers are replaced by few shifters and adders.

Wu et al. have presented a scalable accuracy approximate implementation for exponential function [8]. They used cross-layer optimization to select hardware friendly series approximation and replaced dividers by simple shift operations. These parameters are derived depending on the accuracy requirement.

More recently, Kim et al. have presented template scaling method for the exponential implementation [9]. In this method, they store the template and the difference values in two LUTs and generate the final output by multiplying the two values depending on the input. This is a different presentation of the LUT implementation, wherein the input is divided in two parts and exponential value for each part is looked from the corresponding LUT and multiplied as in [7]. Since, this is based only on LUT, larger LUTs are required for higher precision.

All these methods explore the exponential implementation for both positive and negative domain. As discussed above for most applications, we are interested only in negative domain. A neural network accelerator activation function typically requires only negative domain but requires large number of them for better throughput. So, an optimized circuit for negative domain is quite useful for such applications. This paper explores the optimized hardware implementation of variable precision $e^{-x}$ function.

## II. METHOD OVERVIEW

### A. Principle

The exponential of sum of two numbers can be written as the product of exponential of the individual numbers i.e.

$$e^{x+y} = e^x \times e^y \qquad (2)$$

Let's consider an N-bit binary number $a = b_{N-1}b_{N-2}..b_1b_0$ with $b_0$ being the least significant bit. Then,

$$a = \sum p_i \times b_i \qquad (3)$$

Where $p_i$ is the place value of $i^{th}$ bit. Note that $p_i$ depends on the precision P of the number $a$ and given by $p_i = 2^{-P} \times 2^i$. Now, we can write exponential of $a$ as:

$$e^{-a} = \prod e^{-p_i \times b_i} \qquad (4)$$

Using (4), we can compute the exponential of parts of the input number and multiply them to get the required value. Number of parts can be chosen depending on the choice of LUT depth and number of multipliers. Since, the $e^{-x}$ saturates quite fast, we can divide the input range into two parts:

- **Saturation Region**: For $a \geq 16$, output is saturated to exponential of $(2^{-P} \cdot 16)$, where $P$ is the precision.
- **Non-saturation Region**: Hybrid approach is adopted to compute the $e^{-x}$ in this region. Input a is further divided in two ranges:
  - For values > 1/8, LUTs are used for precise exponential values. This is further divided in two parts: 16 word deep LUT for integer part and 8 word LUT for fractional part.
  - For values ≤ 1/8, Taylor series approximation, which works quite well as can be seen from fig. 1, is used for imprecise exponential computation.

Thus for, non-saturation region, input $a$ is split as given in (5) and exponential is computed as given in (6).

$$a = a_{precise\_1} + a_{precise\_2} + a_{imprecise} \qquad (5)$$

$$e^{-a} = e^{-a_{precise\_1}} \times e^{-a_{precise\_2}} \times e^{-a_{imprecise}} \qquad (6)$$

Where, $a_{precise\_1}$ is the integer part less than 16, $a_{precise\_2}$ is the fractional part > 1/8, and $a_{imprecise}$ is the remainder.

### B. Hardware Oriented Approximations in Series Expansion

The least significant bits, of the fractional part of input $a$, are used for the series approximation circuit. $3^{rd}$ order Taylor series approximation for $e^{-x}$ as given by (7) and re-written as (8).

$$e^{-x} = 1 - x + \frac{x^2}{2!} - \frac{x^3}{3!} + \cdots \qquad (7)$$

$$e^{-x} = 1 - x(1 - \frac{x}{2}(1 - \frac{x}{3})) \qquad (8)$$

Series approximation given in (8) can be approximated for simpler hardware implementation. The last term which requires a divide by 3 operation can be approximated as given in (9), so that it can be implemented using shift and add.

$$e^{-x} = 1 - x(1 - \frac{x}{2}(1 - \frac{2.5x}{8})) \tag{9}$$

The absolute error introduced by this approximation for 16-bit precision is $1.04 \times 10^{-5}$ which is less than one *ulp* (unit in last place or equivalently, bit) as shown in fig. 2.

Note here that Partzsch et al. [7] also implemented the exponential function in similar manner and approximated this divider with a shift and add logic. However, their logic was more complex probably because they opted for higher accuracy. For 16-bit precision, the absolute error in $e^{-x}$ approximation using Partzsch [7] and this method is less than 1 *ulp*. One thing to note here is that their implementation works for both the $e^x$ and $e^{-x}$, so, that may have driven the requirement for higher accuracy. Thus, the idea for this comparison here is only to illustrate that more optimization can be done if we are only interested in $e^{-x}$. However, as we'll see later on, this implementation was adapted only for negative domain for comparison with this work.

## III. HARDWARE IMPLEMENTATION

The implementation has three main blocks; operand splitter, exponential computation for individual parts and final stage of multipliers. The high level block diagram is shown in fig. 3.

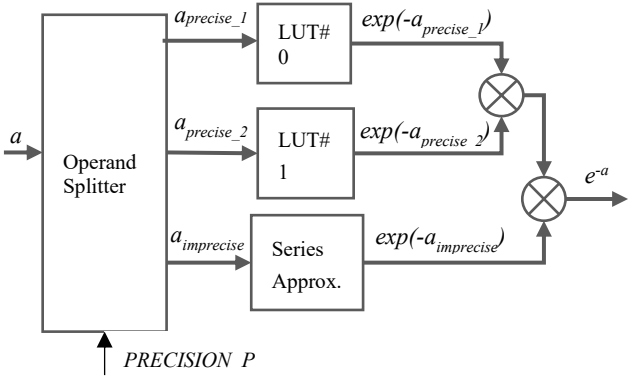

Fig. 3. Top level block diagram of exponential function

### A. Operand Splitter

Operand splitter partitions the input operand depending on the configured precision and range, and sends them to the corresponding computation logic, i.e. LUT or series approximation. Its functioning is explained below in detail.

Let $a = b_N b_{N-1}...b_1 b_0$ be the input binary number and let $P$ be the precision. So, $b_{P-1}...b_1 b_0$ is the fractional part of the number and $b_N b_{N-1}...b_{P-1} b_P$ is the integer part of the number $a$. Then the number $a$ is divided in four parts by the operand splitter as follows:

- Saturation part $a_{sat} = b_N b_{N-1}...b_{P+5} b_{P+4}$
- Integer LUT part $a_{precise\_1} = b_{P+3} b_{P+2} b_{P+1} b_P$
- Fractional LUT part $a_{precise\_2} = b_{P-1} b_{P-2} b_{P-3}$
- Residual part $a_{imprecise} = b_{P-4} b_{P-5} b_{P-6}...b_1 b_0$

The output is saturated if $a_{sat}$ is non-zero. For saturation, maximum value is assigned to the remaining three parts and normal processing is followed. Note that this logic can be implemented as simple combinatorial logic in hardware.

### B. Series Approximation

Though, the series approximation (9) is already optimized by removing a divider; another approximation can be used to further optimize the hardware implementation. The subtractor in (9) is a 2's complement operation as $x$ is a fractional number and subtracting a fractional number from '1' is equivalent to 2's complement operation. For low-cost and imprecise hardware implementation, We can approximate it with 1's complement which is simply bitwise 'not' operation. This simplifies the equation as:

$$e^{-x} = \sim(x * \sim((x \gg 1) * (\sim(x \gg 4 + x \gg 2)))) \tag{10}$$

where $\sim$ is bitwise 'not' operator, and $\gg$ is the right shift operator.

The circuit diagram for series approximation using 1's complement arithmetic is shown in fig. 4. This significantly reduces the hardware cost as the adders between different terms are fully removed now.

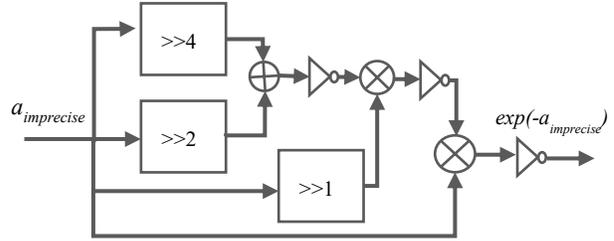

Fig. 4. Circuit for series approximation

### C. Effect of multiplier and LUT precision on accuracy

The accuracy is also affected by the precision of the multipliers and the LUTs, other than the two approximation discussed above. Fig. 5 show the maximum absolute error (MAE), in *ulps*, for some combination of the precision and arithmetic choices. The X and Y axis are used for multiplier precision and MAE respectively. For a given multiplier precision, error is plotted for different LUT precision and arithmetic choices (1's complement or 2's complement) as shown by the legend in the fig. 5. The legend shows LUT precision followed by arithmetic choice; for example, "10,1's" means that LUT precision is 10-bit and 1's complement arithmetic is used for subtractors.

Note that as we increase the precision of multipliers or LUTs, the accuracy improves. The accuracy is dropped when subtractors are approximated by the 1's complement as shown by the inner bars, in lighter color, in the bar chart in fig. 5. However, it is possible to keep the error close to 1 *ulp* with 1's complement circuit. The advantage of 1's complement circuit is that the subtractors are replaced by simple inverters, so logic area is greatly reduced.

### D. Implementation Results

It is evident from the discussion so far that this design requires four multipliers and one adder along with other combinatorial logic. Compared to this design, [7] requires 10 multipliers and four adders. This, reduction in number of multipliers, results in significant improvement of key performance parameters i.e. power, latency and area. For example, this design can compute the function in one cycle against the 6 cycles required by [7]. This comparison may not be totally fair given that original design was intended for the both positive and negative domain and for a greater accuracy. So, the method, proposed in [7], was re-implemented using the same coefficients but reduced number of terms (3 terms instead of four in original paper). Moreover, the 1's complement arithmetic was used along with identical precision and LUT organization. The results were now compared with this modified implementation of [7].

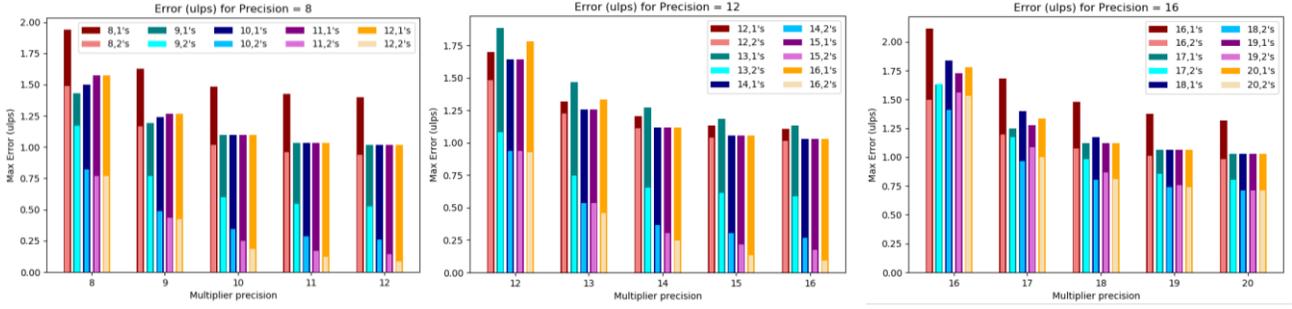

Fig. 5. Impact of multiplier, LUT and arithmetic choices on accuracy for different precision requirements. The figure presents analysis for .8,.12 and .16 output requirement. The horizontal axis shows the precision used for multipliers. The legends represent the LUT precision (first element) and arithmetic choice (2$^{nd}$ element).

This design and modified Partzsch et al. [7] designs were synthesized using 16nm library for the quality of results. The target was 16-bit precision implementation with the error close to 1 *ulp*. The configuration used is 17-bit precision for both multiplier and LUT and 1's complement arithmetic. The area gain for this implementation is 7.48% while the power reduction is 27.72% at 500MHz.

### E. Accuracy of Derived Functions

TABLE I
MAXIMUM ABSOLUTE ERROR FOR DERIVED FUNCTIONS

| Function | Multiplier and LUT precision = 17 | | Multiplier and LUT precision = 19 | |
|---|---|---|---|---|
| | # | Ulps | # | Ulps |
| Gaussian | 2.61e-05 | 1.71 | 1.20e-05 | 0.77 |
| Sigmoid | 2.47e-05 | 1.62 | 5.45e-06 | 0.36 |
| Hyperbolic Tangent | 4.65e-05 | 3.04 | 1.01e-05 | 0.66 |

Table I captures the accuracy of Gaussian and some activation functions derived from e$^{-x}$ implemented using this approximation. For the LUT and multiplier precision of 17 and 1's complement arithmetic, the accuracy loss is less than 2 *ulps* for Gaussian and sigmoid functions while it is close to 3 *ulps* for tanh. However, if we increase the multiplier and LUT precision to 19, the accuracy is within 1 *ulp* for all of them. So, these approximations can be used for such applications.

### IV. VARIABLE WORD-LENGTH IMPLEMENTATION

The implementation discussed above assumes same word length for all the terms of the series expansion. However, the quality of results can be improved by selecting optimal world length for different terms as discussed in this section.

### F. Dependence of Number of terms on Range and Accuracy

It is well known that higher number of terms in series expansion are required to improve the accuracy. More terms are required as the range of the series approximation is increased. Fig. 1 plots the MAE for different ranges for series approximation. Note that the X-axis is the log2 of the range in these plots. Fig. 1a and 1b represent the same data in two different ways for ease of illustration. Fig. 1a presents the maximum absolute error while fig. 1b presents the accuracy in number of fractional bits. Fig. 1b can be directly used to approximate the range for which the results of a particular series approximation are within the fixed point representation used. For example, for $a_{imprecise} < 2^{-8}$, linear, quadratic and cubic approximations are accurate upto 17, 26 and 36 bits respectively after the decimal point.

### G. Variable Precision Mathematics

Since, exponential value for $a_{imprecise}$ is computed by series approximation and rest by look-up tables; the data presented in fig 1 can be used to find the best partitioning for area trade-offs. We can also hypothesize that the additional terms in series approximation are required due to the increased range; so, it should be possible to reduce the precision for higher power terms. So, instead of having fixed world-length for each term, we can have variable word lengths. This will result in the reduction of area and power.

The equation (9) can be partitioned in cubic term $T_c$, square term $T_s$ and linear term $T_l$ as:

$$T_c = 1 - 2.5 \times \frac{x}{8} \tag{11a}$$

$$T_s = 1 - T_c \times \frac{x}{2} \tag{11b}$$

$$T_l = 1 - x \times T_s = e^{-x} \tag{11c}$$

The precision for $T_l$ term is same as that for the output. However, the precision for $T_c$ and $T_s$ can be kept smaller (coarse). The table II presents the accuracy of the approximation for different combinations of the precision for $T_c$ (Cubic term) in rows and $T_s$ (square term) in columns. It can be observed that the accuracy doesn't improve beyond a point by increasing the precision of the cubic term. For example, for 16-bit precision and error requirement within 1 *ulp*, all the configurations in shaded region in Table II are equivalent and 11 and 8-bit precisions are enough for square and cubic terms respectively.

TABLE III
IMPLEMENTATION RESULTS AND COMPARISON WITH STATE OF THE ART

| Implementation | Input Range | Maximum Absolute Error (ulps) | Using SVT cells only (Low Power) | | | Using LVT cells only (High Speed) | | |
|---|---|---|---|---|---|---|---|---|
| | | | Area (um²) | Delay (ps) | Power (nW) | Area (um²) | Delay (ps) | Power (nW) |
| Nilsson et al. [3] | [0-1] | 1 | 1758 | 2528 | 5492 | 1688 | 1686 | 95503 |
| Partzsch et al. [7] | Full (≥0) | 1 | 1567 | 1615 | 4556 | 1597 | 1062 | 88001 |
| Wu et al [8] | [0-1] | 4 | 1335 | 2060 | 4226 | 1277 | 1380 | 74056 |
| This (Fixed WL) | Full (≥0) | 1 | 1543 | 1425 | 4674 | 1520 | 953 | 84474 |
| *This (Variable WL)* | *Full (≥0)* | *1* | *1247* | *1266* | *4060* | *1197* | *855* | *69226* |

TABLE II
ACCURACY OF EXP(-X) APPROXIMATION (IN NUMBER OF BITS AFTER DECIMAL POINT) FOR DIFFERENT PRECISION CHOICES OF CUBIC AND SQUARE TERMS FOR AN INPUT RANGE OF (0-16) AND PRECISION OF $2^{-16}$

| Cubic Term Precision | Square Term Precision | | | | | | |
|---|---|---|---|---|---|---|---|
| | 10 | 11 | 12 | 13 | 14 | 15 | 16 |
| 5 | 13 | 13 | 13 | 13 | 13 | 13 | 13 |
| 6 | 14 | 14 | 14 | 14 | 13 | 13 | 13 |
| 7 | 14 | 14 | 14 | 14 | 14 | 14 | 14 |
| 8 | 14 | 15 | 15 | 14 | 14 | 14 | 14 |
| 9 | 14 | 15 | 15 | 15 | 15 | 15 | 15 |
| 10 | 14 | 15 | 15 | 15 | 15 | 15 | 15 |
| 11 | 14 | 15 | 15 | 15 | 15 | 15 | 15 |
| 12 | 14 | 15 | 15 | 15 | 15 | 15 | 15 |
| 13-16 | 14 | 15 | 15 | 15 | 15 | 15 | 15 |

### H. Hardware Implementation Results

The quality of results for the variable precision implementation is compared with the constant word-length implementation and modified Partzsch [7]. The LUT and linear multipliers precision are kept same for all and only square and cubic term precision is changed to 11 and 8 respectively, for variable word length implementation. The improvement in area and power is 25.83%, 38.62% with respect to constant word length implementation and 31.38% and 55.63% with respect to [7] at 500MHz.

This implementation is also compared with similar implementation published in literature. The published designs were optimized for fair comparison with this implementation and in each case, the computation is done in single cycle. Table III compares the implementation results for best performance (timing) implementation using SVT and LVT cells respectively. It is very clear that the variable word-length implementation achieves the best results.

### V. CONCLUSION

Machine learning and signal processing applications mostly require exponential for negative input range which can be exploited for optimized logic design. This paper presents couple of simple approximations and optimizations that can be used for better implementation such as achieving higher speed or lower power consumption. Such an implementation can be used for designing accelerators for these applications. Furthermore, this circuit can also be used to compute exponential for both positive and negative domain if accuracy requirement is not very high by using a reciprocal unit at the output of this logic.

### REFERENCES

[1] C. Nwankpa, W. Ijomah, A. Gachagan, and S. Marshall, "Activation functions: Comparison of trends in practice and research for deep learning", arXiv preprint, arXiv:1811.03378, 2018

[2] P. T. P. Tang, "Table-lookup algorithms for elementary functions and their error analysis", Proc. of the 10th IEEE Symposium on Computer Arithmetic, pp. 232-236, June 1991, ISBN 0–8186–9151–4

[3] P. Nilsson, A U R Shaik, R Gangarajaiah, and E Hertz, "Hardware implementation of the exponential function using Taylor series", NORCHIP, 2014.

[4] J. E. Volder, "The CORDIC Trigonometric Computing Technique", IRE Transactions on Electronic Computers, vol. EC-8, no. 3, pp. 330-334, 1959.

[5] J. Sudha, M. C. Hanumantharaju, V. Venkateswarulu, and H. Jayalaxmi, "A novel method for computing exponential function using cordic algorithm", Procedia Engineering 30, 2012

[6] P. Pouyan, E. Hertz and P. Nilsson, "A VLSI implementation of logarithmic and exponential functions using a novel parabolic synthesis methodology compared to the CORDIC algorithm", 20th European Conference on Circuit Theory and Design (ECCTD), Linkoping, 2011

[7] J. Partzsch, S. Höppner, M. Eberlein, R. Schüffny, C. Mayr, D. R. Lester, and S. Furber. "A fixed point exponential function accelerator for a neuromorphic many-core system", IEEE International Symposium on Circuits and Systems (ISCAS), 2017.

[8] Di Wu, Tianen Chen, Chienfu Chen, Oghenefego Ahia, Joshua San Miguel, Mikko Lipasti, and Younghyun Kim," SECO: A Scalable Accuracy Approximate Exponential Function Via Cross-Layer Optimization", IEEE/ACM Inter-national Symposium on Low Power Electronics and Design (ISLPED), 2019.

[9] J. Kim, V. Kornijcuk, and D. S. Jeong, "TS-EFA: Resource-efficient high-precision approximation of exponential functions based on template-scaling method", 21st International Symposium on Quality Electronic Design (ISQED), 2020.